\documentclass{superfri}
\usepackage[hidelinks]{hyperref}
\usepackage{algorithm}
\usepackage{algpseudocode}
\usepackage{amsfonts}
\usepackage{amsmath}
\usepackage{amssymb}
\usepackage{anyfontsize}
\usepackage[usenames]{color}
\usepackage{multicol}
\usepackage{wrapfig}
\newcommand{\algrule}[1][.2pt]{\par\vskip.5\baselineskip\hrule height #1\par\vskip.5\baselineskip}
\newcommand*\ceq{\mathrel{\vcenter{\hbox{:}}{=}}}
\newcommand*{\Map}{\mathop{\mathit{Map}}\nolimits }
\newcommand*{\Reduce}{\mathop{\mathit{Reduce}}\nolimits }
\begin{document}

\author{Leonid~B.~Sokolinsky\footnote{\label{susu}South Ural State University (National Research University), Chelyabinsk, Russian Federation}\orcidID{0000-0001-9997-3918} \and  Irina~M.~Sokolinskaya\footnoteref{susu}\orcidID{0000-0002-0717-5378}
}

\title{VaLiPro: Linear Programming Validator for Cluster Computing Systems}

\maketitle{}

\begin{abstract}%
The article presents and evaluates a scalable algorithm for validating solutions to linear programming problems on cluster computing systems. The main idea of the method is to generate a regular set of points (validation set) on a small-radius hypersphere centered at the solution point submitted to validation. The objective function is computed at each point of the validation that belongs to the feasible region. If all the values are less than or equal to the value of the objective function at the point that is to be validated, then this point is the correct solution. The parallel implementation of the VaLiPro algorithm is written in C++ through the parallel BSF-skeleton, which encapsulates all aspects related to the MPI-based parallelization of the program. We provide the results of large-scale computational experiments on a cluster computing system to study the scalability of the VaLiPro algorithm.

\keywords{Linear programming, Solution validator, VaLiPro, Parallel algorithm, Cluster computing system, BSF-skeleton}
\end{abstract}

\section*{Introduction}
\label{sec:intro}
The era of big data~\cite{sokol_1,sokol_2} has generated large-scale linear programming (LP) problems~\cite{sokol_3}. Such problems arise in economics, industry, logistics, statistics, quantum physics, and other fields. To solve them, high-performance computing systems and parallel algorithms are required. Thus, the development of new parallel algorithms for solving LP problems and the revision of current algorithms have become imperative. As examples, we can cite the works~\cite{sokol_4, sokol_5,sokol_6,sokol_7,sokol_8,sokol_9}. The development of new parallel algorithms for solving large-scale linear programming problems involves testing them on various benchmarks. One of the most well-known benchmark repositories of linear programming problems is the Netlib-Lp benchmark suite~\cite{sokol_10}. The solutions to all the problems from this repository are known. At the same time, in practice, it is often necessary to test a new algorithm on certain problems with unknown solutions. When testing an LP solver on such classes of problems, there is a need for validation (certification) and refinement of the obtained solution.

Several works have been devoted to the problem of certification and refinement of LP solutions. The paper~\cite{sokol_11} presents the LPlex system, which verifies and repairs a given solution to an LP problem for feasibility and optimality using exact arithmetic to guarantee the correctness of the results. The LPlex system can solve medium to large LP problems to optimality. Based on exact arithmetic (integer, rational, or modular), LPlex implements a module to detect block structures in matrices~\cite{sokol_12} and supports LU-factorizations of sparse matrices, the Bareiss method \cite{sokol_13,sokol_14}, and the Wiedemann method~\cite{sokol_15}. The main drawback of the approach is that LPlex fails if the certified solution is not close enough to the optimal one. Koch~\cite{sokol_16} modified this approach to computing optimal solutions for the full set of Netlib-Lp instances. Rather than attempting to repair a nonoptimal basis with rational pivots, Koch recomputes a floating-point solution using greater precision in the floating-point representations. He employed the long double type that specifies 128-bit values. In~\cite{sokol_17}, Applegate and co-authors extend Koch's methodology with an implementation that dynamically increases the precision of floating-point computations until a rational solution satisfying the optimality condition is obtained. They modify the conventional simplex algorithm by changing every floating-point type into the rational type provided by the GNU multiple precision arithmetic library (GMP) \cite{sokol_18} and replacing every arithmetic operation in the original code with the corresponding GMP operations. The program starts with the best native floating-point precision and then increases it by about 50\% at each iteration (keeping the precision value a multiple of 32 bits to align with the typical word size). The main drawback of this approach is that the use of the multiple-precision arithmetic in the case of large and complex LP problems has high overheads. In~\cite{sokol_19}, Panyukov and Gorbik try to overcome this disadvantage by using parallel computing on distributed memory. In this paper, they utilize rational arithmetic and propose two approaches for parallelizing the simplex method. The first method is based on the decomposition of the simplex tableau by columns. The second method is based on the modified simplex method using the inverse matrix and exploits the decomposition of the original matrix by columns and that of the inverse matrix by rows. However, the results of computational experiments are not sufficiently convincing for the following reasons: there is no comparison with the best sequential solutions; the computations were performed using only three sparse LP problems (the number of nonzero elements did not exceed 5\%); and, in addition, the scalability bound of the proposed parallel algorithm was only 16 processors. Another original approach is suggested by Gleixner and co-authors in~\cite{sokol_20}. This paper describes an iterative refinement procedure for computing extended-precision or exact solutions to LP problems. Arbitrarily precise solutions can be computed by solving a sequence of closely related LPs with limited-precision arithmetic. These LPs share the same constraint matrix as the original problem instance and are transformed only by modification of the objective function, right-hand sides, and variable bounds. This implementation is publicly available as an extension of the academic LP solver SoPlex.

All the methods discussed above concentrate on refining the approximate solution that has already been found. If the found solution is too far from the correct one, which means that there is an error in the algorithm, then the use of these methods becomes impractical. In addition, all of these algorithms have high computational complexity and do not allow efficient parallelization on large cluster computing systems. The method proposed in this article focuses on debugging and validating new LP algorithms on cluster computing systems. It is implemented as a parallel program, VaLiPro (Validator of Linear Program), which shows good scalability on multiprocessor computing systems with distributed memory. The rest of the paper is organized as follows. Section~\ref{sokol_Method} provides a formal description of the proposed method for validating solutions to LP problems and presents a sequential version of the VaLiPro algorithm. The parallel version of the VaLiPro algorithm is discussed in Section~\ref{sokol_Parallel_algorithm}. Section~\ref{sokol_Implementation} describes the implementation of the VaLiPro parallel algorithm in C++ using the BSF-skeleton. Here, we present the results of computational experiments on a cluster computing system, which confirm the efficiency of the proposed approach. In Section~\ref{sokol_Conclusion}, we summarize the obtained results and expose plans for using the VaLiPro validator in the development of an artificial neural network capable of solving large-scale LP problems.

\section{Method for Validating Solutions to LP Problems}\label{sokol_Method}
Let the following linear programming problem be given in the Euclidean space ${\mathbb{R}^n}$:
\begin{equation}\label{sokol_Formula1}\bar x = \arg \max \left\{ {\left\langle {c,x} \right\rangle \mid {Ax \leqslant b,x \in {\mathbb{R}^n}} } \right\},
\end{equation}
\begin{wrapfigure}{r}{0.5\linewidth}
  \centering
  \includegraphics[scale=1]{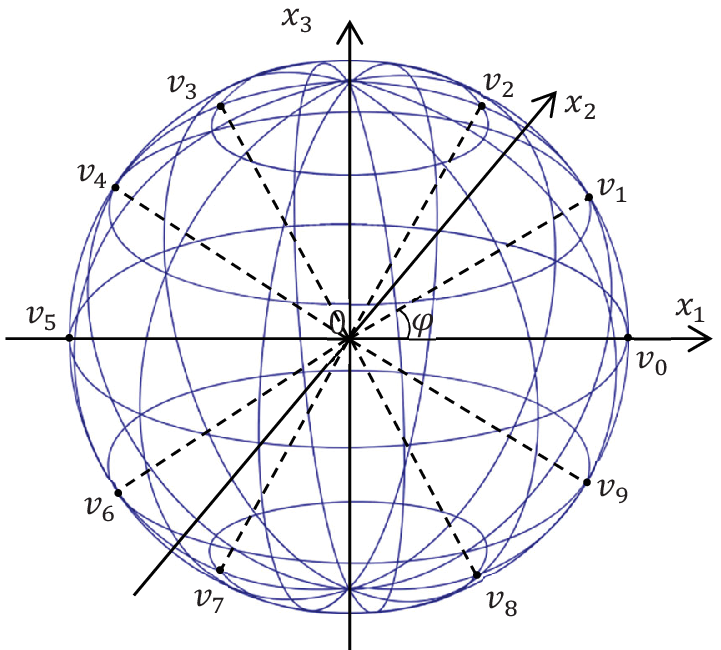}
  \caption{Plotting the points of the validation set $V$ on a three-dimensional sphere with $d=5$}
  \label{sokol_Fig1}
\end{wrapfigure}
where $c$ is the vector of the objective function coefficients. Here and below, $\langle{\cdot, \cdot}\rangle $ stands for the dot product of vectors. Let us define $M = \left\{ {x \in {\mathbb{R}^n} \mid {Ax \leqslant b}} \right\}$ as the feasible region of problem~\eqref{sokol_Formula1}. By definition, the set $M$ is convex and closed. From now on, we assume that $M$ is a nonempty bounded set, i.e., problem~\eqref{sokol_Formula1} has at least one solution. Let $\tilde x \in {\mathbb{R}^n}$ be an approximate solution of problem~\eqref{sokol_Formula1} obtained using some LP solver that must be certified.

The main idea of the VaLiPro validation method is to construct a finite set of points~$V$ covering a hypersphere $S$ of small radius $\rho$ centered at the certified solution point~$\tilde x$:
\[V \subset S = \bigl\{ x \in \mathbb{R}^n \bigm\vert {{\left\| {x - \tilde x} \right\|}^2 = \rho ^2} \bigr\}.\]
Here and below, $\left\|  {}\cdot{}  \right\|$ denotes the Euclidean norm. Let us compute the maximum of the objective function on the set $V\cap M$:
\[\bar v = \arg \max \left\{\left\langle {c,v} \right\rangle \bigm\vert v \in V \cap M \right\}.\]
If $\bigl\lvert {\left\langle {c,\bar v} \right\rangle  - \left\langle {c,\tilde x} \right\rangle } \bigr\rvert < \varepsilon$, then the approximation $\tilde x$ is considered correct. Otherwise, $\tilde x$ is considered an incorrect solution. Here, $\varepsilon\in{\mathbb{R}_{>0}}$ is a small positive constant that is a parameter of the validation algorithm.

Let us describe the method for constructing the validation set $V$. It is known~\cite{sokol_21} that the coordinates of any point $v = ({v_1}, \ldots ,{v_n})$ lying on the surface of the hypersphere $S$ defined by the equation
\[{\left\| {x - \tilde x} \right\|^2} = {\rho ^2}\]
can be represented as follows:
\begin{equation}\label{sokol_Formula2}
\begin{aligned}   {v_1} &= \rho \cos(\phi _1);  \\
{v_j} &= \rho \cos(\phi _j)\prod\limits_{i = 1}^{j - 1} {\sin(\phi _i)} \; (j = 2, \ldots ,n - 2); \\
{v_{n - 1}} &= \rho \sin(\theta) \prod\limits_{i = 1}^{n - 2} {\sin(\phi _i)}; \\
{v_n} &= \rho \cos(\theta) \prod\limits_{i = 1}^{n - 2} {\sin(\phi _i)},
\end{aligned}
\end{equation}
where $0 \leqslant {\phi _j} \leqslant \pi $ ($j = 1, \ldots ,n - 2$) and $0 \leqslant \theta  < 2\pi $. Let us explain the method for generating the validation set $V$ using a three-dimensional sphere (see Figure~\ref{sokol_Fig1}). Fix an odd number of \emph{parallels} $d \geqslant 3$ (poles are excluded). Set
\begin{equation}\label{sokol_Formula3}\varphi  = \pi /d.\end{equation}
In the plane $({x_1},0,{x_3})$, we set the angles $0,\varphi ,\ldots,(2d - 1)\varphi $ starting from the axis $(0,x)$. At the intersection with the sphere, the resulting rays give the set of points $\left\{ {{v_0}, \ldots, {v_{2d - 1}}} \right\}$, which uniquely define $d$ parallels. Then, on the plane $({x_1}, 0, {x_2})$, set the angles $0,\varphi ,\ldots,(2d-1)\varphi $ from the axis $(0, {x_1})$ and define $d$ \emph{meridians} in the same way. The intersections of parallels and meridians, excluding the poles, give the points that form a validation set on the three-dimensional sphere.

The described method for generating points of the validation set for $n \geqslant 3$ in the general case is given in Algorithm~\ref{sokol_alg1}a.
\begin{algorithm}[t]
\caption{Generating points of the validation set}\label{sokol_alg1}
\textbf{Parameters}: $d,\rho$
\algrule
\begin{multicols}{2}
\begin{center}
\textbf{a) With duplicates} \\
\textbf{b) Without duplicates}
\end{center}
\end{multicols}
\algrule
\begin{multicols}{2}
\begin{algorithmic}[1]
\State $\varphi\ceq \pi/d$
\For{$j_{n-1}=0\ldots\left(2d-1\right)$}
\State $\theta\ceq j_{n-1}\varphi$
\For{$j_{n-2}=0\ldots d$}
\State $\phi_{n-2}\ceq j_{\left(n-2\right)}\varphi$
\State $\ldots$
\For{$j_2=0\ldots d$}
\State $\phi_2\ceq j_2\varphi$
\For{$j_1=0\ldots d$}
\State $\phi_1\ceq j_1\varphi$
\State $\varpi\ceq 1$
\State $v_1\ceq \rho \cos(\phi_1)$
\For{$l=2\ldots n-2$}
\State $\varpi\ceq\sin(\phi_{l-1})\varpi$
\State $v_l\ceq \rho\cos(\phi_l)\varpi$
\EndFor
\State $v_{n-1}\ceq \rho\sin(\theta)\varpi$
\State $v_n\ceq \rho\cos(\theta)\varpi$
\State \textbf{output} $v$
\EndFor
\EndFor
\State \ldots
\EndFor
\EndFor
\State \textbf{stop}
\end{algorithmic}
\begin{algorithmic}[1]
\State $\varphi\ceq \pi/d$
\For{$j_{n-1}=0\ldots\left(2d-1\right)$}
\State $\theta\ceq j_{n-1}\varphi$
\For{\textcolor{red}{$j_{n-2}=1\ldots d-1$}}
\State $\phi_{n-2}\ceq j_{\left(n-2\right)}\varphi$
\State $\ldots$
\For{\textcolor{red}{$j_2=1\ldots d-1$}}
\State $\phi_2\ceq j_2\varphi$
\For{\textcolor{red}{$j_1=1\ldots d-1$}}
\State $\phi_1\ceq j_1\varphi$
\State $\varpi\ceq 1$
\State $v_1\ceq \rho \cos(\phi_1)$
\For{$l=2\ldots n-2$}
\State $\varpi\ceq\sin(\phi_{l-1})\varpi$
\State $v_l\ceq \rho\cos(\phi_l)\varpi$
\EndFor
\State $v_{n-1}\ceq \rho\sin(\theta)\varpi$
\State $v_n\ceq \rho\cos(\theta)\varpi$
\State \textbf{output} $v$
\EndFor
\EndFor
\State \ldots
\EndFor
\EndFor
\State \textbf{stop}
\end{algorithmic}
\end{multicols}
\end{algorithm}
The nested loops with the headers in steps 2, 4,\,\dots, 7, and 9 generate the following spherical coordinates of a validation point:
\begin{equation}\label{sokol_Formula4}
\left( {\rho ,{\phi _1},{\phi _2}, \ldots ,{\phi _{n - 2}},\theta } \right).
\end{equation}
\begin{wrapfigure}{l}{0.4\linewidth}
    \begin{minipage}{0.37\textwidth}
        \begin{algorithm}[H]\caption{Function $g$ (calculating the point $v$ by its number $k$)}\label{sokol_alg2}
        \begin{algorithmic}[1]
            \Function {$g$}{$k,d,\rho$}
            \State $u_{n-1} \ceq\ \left\lfloor k/\left(d-1\right)^{n-2}\right\rfloor$
            \State $u_n \ceq u_{n-1}$
            \State $k\ceq k\mod (d-1)^{n-2}$
            \For {$j=\left(n-3\right)\ldots0$}
            \State $u_j\ceq \left\lfloor k/(d-1)^j\right\rfloor+1$
            \State $k\ceq k\mod(d-1)^j$
            \EndFor
            \State $\varpi\ceq 1$
            \State $\varphi\ceq \pi/d$
            \State $v_1\ceq \rho\cos(u_1\varphi)$
            \For {$j=2\ldots\left(n-2\right)$}
            \State $\varpi\ceq\varpi\sin(u_{j-1}\varphi)$
            \State $v_j\ceq\rho\cos(u_j\varphi)\varpi$
            \EndFor
            \State $\varpi\ceq\varpi\sin(u_{n-2}\varphi)$
            \State $v_{n-1}\ceq \rho\sin(u_{n-1}\varphi)\varpi$
            \State $v_n \ceq \rho\ \cos(u_n\varphi)\varpi$
            \State \Return $(v_1,\dots,v_n)$
            \EndFunction
            \end{algorithmic}
        \end{algorithm}
    \end{minipage}
\end{wrapfigure}
In steps~11--18, the spherical coordinates are converted to Cartesian coordinates by Equations~\eqref{sokol_Formula2}. Multiplying the quantities of iterations of for-loops with headers 2, 4,\,\dots, 7, and 9, we conclude that Algorithm~\ref{sokol_alg1}a outputs $2d{(d + 1)^{n - 2}}$ validation points. However, there will be duplicates among the output points. The computational experiments showed that if one sets the dimension $n = 4$ and the number of parallels $d = 5$, then Algorithm~\ref{sokol_alg1}a generates 189 duplicates with a total number of points equal to 360, which is more than 50\%. The duplicates are generated at iterations in which ${\phi _i} = 0$ or ${\phi _i} = \pi$, which corresponds to ${j_i} = 0$ and ${j_i} = d$ ($i = 1, \ldots ,n - 2$). The reason is that one of the factors $\sin(\phi _i)$ in~\eqref{sokol_Formula2} is equal to zero in this case, and therefore the variations of other factors cannot change the value of the corresponding coordinate. This issue can be solved without a major revision of Algorithm~\ref{sokol_alg1}a, by changing the start values and end values of the control variables in loop headers 4,\,\dots, 7, and 9, as is done in Algorithm~\ref{sokol_alg1}b. This algorithm generates a validation set without duplicates but, at the same time, it loses a certain number of unique points. With $n = 4$ and $d = 5$, this quantity is 11, which is less than 7\% of the whole set after removing duplicates. Experiments have shown that such a loss does not significantly affect the accuracy of the validation algorithm. The number of points of the validation set $V$ generated by Algorithm~\ref{sokol_alg1}b is determined by the following equation:
\begin{equation}\label{sokol_Formula5}\lvert V \rvert = 2d(d - 1)^{n - 2}.\end{equation}
The main drawback of Algorithm~\ref{sokol_alg1}b is that the number of nested loops depends on the problem dimension, which does not allow using the dimension as a program parameter. To overcome this drawback, we use a vector-valued function that calculates the coordinates of a point by its sequential number $k$ in the point sequence generated by Algorithm~\ref{sokol_alg1}b (counting starts from zero). The definition of this function is given in Algorithm~\ref{sokol_alg2}.

The final implementation of the VaLiPro method, using the vector-valued function $g$, is given in Algorithm~\ref{sokol_alg3}. An additional parameter of this algorithm is the small positive constant $\varepsilon$ (by default, $\varepsilon = {10^{ - 6}}$), which compensates for possible numerical errors when comparing the values of the objective function in Step~5. Let us make several brief comments on the steps of Algorithm~\ref{sokol_alg3}. Step~1 reads the source data of LP problem~\eqref{sokol_Formula1}, the algorithm parameters, and the solution~$\tilde x$ that is to be certified. Step~2 calculates the angle $\varphi$ according to Equation~\eqref{sokol_Formula3}. Step~3 begins the loop that varies the point number $k$ from $0$ to $2d{(d - 1)^{n - 2}} - 1$ as per Equation~\eqref{sokol_Formula5}. Using the vector-valued function $g$ (see Algorithm~\ref{sokol_alg2}), Step~4 computes the next validation point~$v$. Step~5 checks whether~$v$ belongs to the feasible region of problem~\eqref{sokol_Formula1} and compares the objective-function values at the points~$v$ and~$\tilde x$. If the objective function takes a larger value at the point~$v$, and this point is feasible, then the control is passed to Step~9, which prints a message stating that the certified solution is not correct. Otherwise, the next iteration of the loop proceeds. If the loop ends naturally, the control is passed to Step~7, which outputs a message saying that the solution is correct. After that, the control is passed to Step~10, which completes the execution of the algorithm.

\begin{figure}[t]
\begin{center}
\begin{minipage}{0.5\textwidth}
     \begin{algorithm}[H]\caption{Validation of the LP solution $\widetilde{x}$}\label{sokol_alg3}
        \begin{algorithmic}[1]
            \State \textbf{input} $n,A,b,c,d,\rho,\varepsilon,\widetilde{x}$
            \State $\varphi \ceq \pi/d$
            \For {$k=0\ldots 2d(d-1)^{n-2}-1$}
            \State $v \ceq g(k,d,\rho)$
            \State \textbf{if} $Av \leqslant b \And \langle c,v\rangle  > \langle c,\widetilde{x}\rangle + \varepsilon$ \textbf{goto} 9
            \EndFor
            \State \textbf{output} ``Solution is correct''
            \State \textbf{goto} 10
            \State \textbf{output} ``Solution is incorrect''
            \State \textbf{stop}
        \end{algorithmic}
     \end{algorithm}
\end{minipage}
\end{center}
\end{figure}

\section{Parallel Algorithm for Validating LP Solutions}\label{sokol_Parallel_algorithm}
According to Equation~\eqref{sokol_Formula5}, the cardinality of the validation set generated by Algorithm~\ref{sokol_alg3} depends exponentially on the space dimension. Therefore, Algorithm~\ref{sokol_alg3} has high computational complexity for large dimensions. To reduce computational overheads, we developed a parallel version of Algorithm~\ref{sokol_alg3}, given as Algorithm~\ref{sokol_alg4} below. The parallel algorithm is based on the BSF parallel computation model~\cite{sokol_22,sokol_23}, which exploits the master--slave paradigm~\cite{sokol_24}. According to the BSF model, the master node serves as a control and communication center. All slave nodes execute the same code but on different data. The BSF model assumes the algorithm representation in the form of operations on lists using the higher-order functions $\Map$ and $\Reduce$ defined by the Bird--Meertens formalism~\cite{sokol_25}. The higher-order function $\Map$ transforms the original list $W = \left[ {{w_0}, \ldots ,{w_{K - 1}}} \right]$ into the list $Z = \left[ {{z_0}, \ldots ,{z_{K - 1}}} \right]$ by applying the function ${f_{\tilde x}}$ to each element:
\[Z = \Map({f_{\tilde x}},W) = \left[ {{f_{\tilde x}}({w_0}), \ldots ,{f_{\tilde x}}({w_{K - 1}})} \right].\]
In the case considered here, the elements of the list $W$ are the sequential numbers of the validation set points, that is,
\[W = \left[ {0, \ldots ,K - 1} \right],\]
where $K = 2d(d - 1)^{n - 2}$. The Boolean function ${f_{\tilde x}}:\{ 0, \ldots ,K - 1\}  \to \{ \mathit{true},\mathit{false}\} $ is defined as follows:
\[{f_{\tilde x}}(w) = \left\{
\begin{aligned}
  \mathit{true} &   {{}\bigm\vert {A \cdot g(w) \leqslant b \wedge \left\langle {c,g(w)} \right\rangle  \leqslant \left\langle {c,\tilde x} \right\rangle ;} } \\
  \mathit{false} & {{}\bigm\vert {A \cdot g(w) > b \vee \left\langle {c,g(w)} \right\rangle  > \left\langle {c,\tilde x} \right\rangle ,} }
\end{aligned} \right.\]
where the vector-valued function~$g$ computes the coordinates of the validation point by its number~$w$. The function~${f_{\tilde x}}$ returns $\mathit{true}$ if the point~$g(w)$ belongs to the feasible region and if the value of the objective function at this point is less than or equal to the value of the objective function at the point~$\tilde x$. Otherwise, the function~${f_{\tilde x}}$ returns $\mathit{false}$. Thus, the list $Z = \left[ {{z_0}, \ldots ,{z_{K - 1}}} \right]$ contains Boolean indicators for all points of the validation set. If at least one element in this list has the value $\mathit{false}$, then the point~$\tilde x$ is an incorrect solution of problem~\eqref{sokol_Formula1}.
\begin{algorithm}[t]
\caption{Parallel algorithm for validating an LP solution}\label{sokol_alg4}
\begin{multicols}{2}
\begin{center}
\textbf{Master} \\
\textbf{Slave (\emph{l}=0,\dots,\emph{L-1})}
\end{center}
\end{multicols}
\algrule
\begin{multicols}{2}
\begin{algorithmic}[1]
\State
\State
\State
\State
\State
\State
\State \textbf{RecvFromSlaves} $[s_0,\ldots,s_{L-1}]$
\State $s\ceq \Reduce(\land,[s_0,\ldots,s_{L-1}])$
\If {$s=true$}
\State \textbf{output} ``Solution is correct''
\Else
\State \textbf{output} ``Solution is incorrect''
\EndIf
\State \textbf{stop}
\end{algorithmic}
\begin{algorithmic}[1]
\State \textbf{input} $n,A,b,c,d,\rho,\varepsilon,\widetilde{x}$
\State $L\ceq \textbf{NumberOfSlaves}$
\State $K\ceq 2d(d-1)^{n-2}$
\State $W_l\ceq \left[lK/L,\ldots,(l+1)K/L-1\right]$
\State $Z_l \ceq \Map\left(f_{\widetilde{x}},W_l\right)$
\State $s_l\ceq \Reduce(\land,Z_l)$
\State \textbf{SendToMaster} $s_l$
\State
\State
\State
\State
\State
\State
\State \textbf{stop}
\end{algorithmic}
\end{multicols}
\end{algorithm}

The higher-order function $\Reduce$ transforms the list $Z = \left[ {{z_0}, \ldots ,{z_{K - 1}}} \right]$ into a single Boolean value $s$ by iteratively applying the conjunction operation to all the elements of the list $Z$:
\[s = \Reduce(\wedge ,Z) = {z_0} \wedge  \ldots  \wedge {z_{K - 1}}.\]

In Step~4 of Algorithm~\ref{sokol_alg4}, the $l$-th slave sets its own part $W_l$ of the list $W$:
\[{W_l} = \left[ {lK/L, \ldots ,(l + 1)K/L - 1} \right].\]
Here, $L$ denotes the number of slaves. For simplicity, we assume that $K$ is a multiple of $L$. In Step~5, the slave applies the $\Map$ function to its sublist $W_l$. In Step~6, the resulting sublist of Boolean values is folded into a single Boolean value ${s_l}$ by applying the $\Reduce$ function, taking the conjunction operation as the first parameter. In Step~7, the $l$-th slave sends the value $s_l$ to the master. In the same Step~7, the master receives all the calculated values from the slaves. In Step~8, the master folds the list of received values into a single Boolean value $s$ using the $\Reduce$ function. In steps~9--12, the master examines the calculated Boolean value $s$ and outputs the corresponding conclusion.

\section{Software Implementation and Computational Experiments}\label{sokol_Implementation}

\begin{figure}[t]
  \centering
  \includegraphics[scale=1]{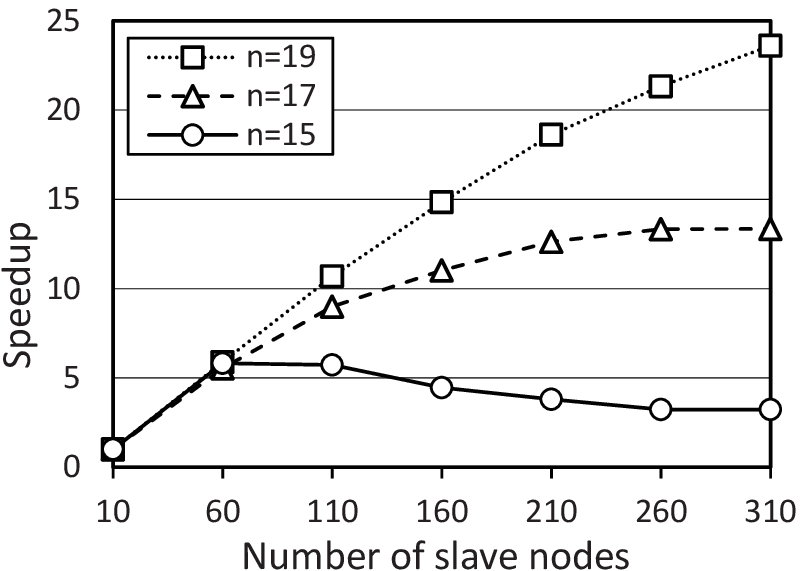}
  \caption{Speedup curves of the VaLiPro parallel algorithm for various dimensions}
  \label{sokol_Fig2}
\end{figure}

We implemented the parallel Algorithm~\ref{sokol_alg4} in C++ through the parallel BSF-skeleton~\cite{sokol_26}, which is based on the BSF parallel computation model~\cite{sokol_22,sokol_23} and encapsulates all aspects related to the parallelization of the program using the MPI library~\cite{sokol_27}. The source code of the VaLiPro parallel program is freely available at \url{https://github.com/leonid-sokolinsky/BSF-LPP-Validator}. Using this program, we conducted large-scale computational experiments on the cluster computing system ``Tornado SUSU''~\cite{sokol_28}. The specifications of this system are given in Table~\ref{sokol_Table1}.
\begin{table}[t]
\caption{Specifications of the ``Tornado SUSU'' computing cluster}
\centering
\begin{tabular}{l|l}
  \hline
  Parameter & Value \\
  \hline
  Number of processor nodes & 480 \\
  Processor & Intel Xeon X5680 (6 cores, 3.33 GHz) \\
  Processors per node & 2\\
  Memory per node & 24 GB DDR3\\
  Interconnect & InfiniBand QDR (40 Gbit/s) \\
  Operating system & Linux CentOS\\
  \hline
\end{tabular}\label{sokol_Table1}
\end{table}
For experiments, we used random LP problems generated by the program FRaGenLP~\cite{sokol_29} with the following parameters: $\alpha  = 200$ (the length of the bounding-hypercube edge), $\theta  = 100$ (the radius of the large hypersphere), $\rho  = 50$ (the radius of the small hypersphere),  ${L_{\mathrm{max}}} = 0.35$ (the upper bound of \emph{near parallelism} for hyperplanes), ${S_{\mathrm{min}}} = 100$ (the minimum acceptable \emph{closeness} for hyperplanes), ${a_{\mathrm{max}}} = 1000$ (the upper absolute bound for the coefficients),  and ${b_{\mathrm{max}}} = 10\,000$ (the upper absolute bound for the constant terms). The experiments were conducted for the following dimensions: $n = 15$, $n = 17$, and $n = 19$. The numbers of inequalities were $46$, $52$, and $58$, respectively. The solutions to the LP problems were obtained using the apex method~\cite{sokol_8}. Throughout the experiments, we used the following VaLiPro parameters: $d = 5$, $\rho  = 1$, and $\varepsilon  = {10^{ - 6}}$. The results of the experiments are shown in Figure~\ref{sokol_Fig2}. The verification of a solution for a problem of dimension $n = 19$ with a configuration consisting of the master node and one slave node took 17~minutes. The verification of a solution for the same problem with a configuration consisting of the master node and $310$ slave nodes took 4~seconds. The analysis of the results showed that the scalability bound (the maximum of the speedup curve) of the algorithm significantly depends on the dimension of the problem. For $n = 19$, the parallel version of the VaLiPro algorithm demonstrated near-linear scalability up to 310 processor nodes. For $n = 17$, the scalability bound was approximately 260 nodes, and for $n = 15$, this bound decreased to 60 processor nodes. This is because a problem of such a small dimension is not able to load such a large number of processor nodes: the time spent on data transfer over the network begins to dominate over the time spent on calculations, and the processors begin to stand idle.

\section{Conclusions}\label{sokol_Conclusion}

The article presents the parallel algorithm VaLiPro for validating linear programming solutions on cluster computing systems. The main idea of the validation algorithm is to generate a regular set of points on a small-radius hypersphere centered at the solution point that is to be certified. The solution is considered correct if all points of the validation set belonging to the feasible region have lower values of the objective function than does the solution point being certified. The implementation of the parallel algorithm VaLiPro was performed in C++ using the parallel BSF-skeleton, which encapsulates in the problem-independent part of its code all aspects related to the parallelization of a program using the MPI library. The source code of the developed parallel program is freely available at \url{https://github.com/leonid-sokolinsky/BSF-LPP-Validator}. The proposed validation method is generic and suitable for linear programming problems of any kind. The advantage of the parallel VaLiPro algorithm is the near-linear speedup starting with a problem dimension of 19. The main drawback that limits the practical use of the suggested method is the exponential growth of the number of points in the validation set as the dimension of the space increases; this results in the exponential growth of the computational complexity. The described algorithm was used together with the FRaGenLP generator and the apex method to prepare a training dataset of 70\,000 examples, which will be used to develop an artificial neural network capable of solving multidimensional linear programming problems.

\section*{Acknowledgements}
The reported study was partially funded by the Russian Foundation for Basic Research (project No.~20-07-00092-a) and the Ministry of Science and Higher Education of the Russian Federation (government order FENU-2020-0022).

\openaccess

\bibliographystyle{superfri}


\end{document}